\documentclass[10pt]{iopart}
\usepackage[dvips]{graphicx}
\usepackage{color,epsfig}
\newcommand{\ir}{\textrm{i}}
\newcommand{\dr}{\textrm{d}}

\begin{document}
\title[Diffractive orbits in a 2D microwave cavity]{Diffractive orbits in the length spectrum of a 2D microwave cavity with a small scatterer}
\author{David Laurent, Olivier Legrand, Fabrice Mortessagne}
\address{Laboratoire de Physique de la Mati\`ere Condens\'ee, CNRS UMR 6622,\\  Universit\'e de Nice-Sophia Antipolis, 06108 Nice, France}
\ead{david.laurent@unice.fr, olivier.legrand@unice.fr, fabrice.mortessagne@unice.fr}
\maketitle

\begin{abstract}
In a 2D rectangular microwave cavity dressed with one point-like scatterer, a semiclassical approach is used to analyze the spectrum in terms of periodic orbits and diffractive orbits. We show, both numerically and experimentally, how the latter can be accounted for in the so-called \emph{length spectrum} which is retrieved from 2-point correlations of a finite range frequency spectrum. Beyond its fundamental interest, this first experimental evidence of the role played by diffractive orbits in the spectrum of an actual cavity, can be the first step towards a novel technique to detect and track small defects in wave cavities.
\end{abstract}
\maketitle

\section{Introduction}

It has been widely recognized that the semiclassical approach of spectral statistics based on periodic orbits (POs) has met a definite success in chaotic systems (see e.g. \cite{Stoeckmann99} and references therein). In systems where one or more point-like scatterers are added, a similar approach is possible, calling for both POs and the so-called diffractive orbits (DOs) \cite{Vattay94}. While the proliferation with length of diffractive orbits may be reminiscent of the proliferation of POs in chaotic systems, if the unperturbed system is regular, the statistics is clearly not predicted by Random Matrix Theory even if some level repulsion appears due to the presence of scatterers. Indeed, at large spacing, the spacing distribution decays exponentially like in a regular system \cite{BogGirSch2002,BogGerSch2001}. In fully chaotic systems, on the contrary, adding scatterers has been shown to be of practically no effect on the spectral statistics \cite{BogLebSch2000}.

In the present paper, we will show both numerically and experimentally, how the presence of short DOs can be accounted for in the so-called \emph{length spectrum} which one can retrieve from two-point correlations of a finite range frequency spectrum. Our experiments are performed in microwave cavities at room temperature and
the limited range of the available spectrum is a practical issue.
Nevertheless, in the case of a  2D rectangular (regular) cavity with one point-scatterer (actually a small scatterer in the experiments), the lengths associated to the shortest DOs are easily identified in our experiments and can be used to locate the scatterer and possibly track its displacements.

\section{Semiclassical expansion and spectral density}

We now briefly recall the semiclassical approach of spectral correlations based on a decomposition of the spectral density on POs. It relies on the semiclassical representation of the Green's function $G( \vec{r}, \vec{r}^{\,\prime})$, written as a sum over classical trajectories connecting points $\vec{r}$ and $\vec{r}^{\,\prime}$ 

\begin{equation}\label{Green}
G( \vec{r}, \vec{r}^{\,\prime}) = \sum_{cl. tr.} A_{tr} (\vec{r}, \vec{r}^{\,\prime}) \exp \left( \ir \mathcal{S}_{tr} (\vec{r}, \vec{r}^{\,\prime}) \right)
\end{equation}
By evaluating the trace of $G(\vec{r},\vec{r})$ through a stationary phase approximation, one obtains a semiclassical trace formula for the modal density $\rho$ \cite{Stoeckmann99}. The latter is thus written as an average smooth density plus an oscillatory part:

\begin{equation}
\label{rho}
\rho(k) = \overline{\rho}(k) + \rho^{osc}(k)
        = \overline{\rho}(k) + \sum_{j} A_{j} e^{\ir k \ell_{j}} + \textrm{c.c.}
\end{equation}
Here, $k$ is the wavenumber, $\ell_j$ is the length of the $j$th PO and $A_j$ is its complex amplitude accounting for its stability and possibly depending on $k$.

In diffractive systems with point-like singularities, classical trajectories that hit those singularities can be continued in any direction. These can nonetheless be tackled with in the wave description by introducing an isotropic diffraction coefficient $\mathcal{D}$, which fixes the scattering amplitude at each scatterer. 
In previous works (see \cite{ExnerSeba96,RahavFishman2002}), this diffraction constant has been calculated (with the free Green's function in \cite{RahavFishman2002}) to yield:
\begin{equation}\label{DiffConst}
\mathcal{D} = \frac{2 \pi}{-\ln (ka/2) - \gamma + \ir \pi / 2}
\end{equation}
where $\gamma$ is the Euler constant and $a$ is a characteristic length which may be interpreted as the non-vanishing radius of an s-wave scattering disc  \cite{ExnerSeba96}, the above expression being precisely the limiting scattering amplitude for $k a \ll 1$.
The semiclassical expansion of the Green's function in the presence of a point scatterer located at $\vec{s}$ therefore reads:
\begin{equation}\label{GreenScat}
G( \vec{r}, \vec{r}^{\,\prime}) = G_0( \vec{r}, \vec{r}^{\,\prime}) + G_0( \vec{r}, \vec{s} )\,\mathcal{D}\, G( \vec{s}, \vec{r}^{\,\prime} )
\end{equation}
where $G_0$ is the unperturbed Green's function. Hence formula (\ref{rho}) for the modal density still holds provided that nonclassical contributions due to DOs are included, consistently with the geometrical theory of diffraction, yielding: $\rho^{osc}=\rho^{osc}_{po}+\rho^{osc}_{do}$.  In a rectangular domain of area $\mathcal{A}$ with a single point scatterer, contributions from periodic orbits and diffractive orbits respectively read \cite{Pavloff1995}:
\begin{equation}
\label{po}
\rho^{osc}_{po}(k)=\frac{\mathcal{A}}{\pi}{\sum_{po}}^\prime\sum_{r=1}^\infty\frac{k}{\sqrt{2\pi r k\ell_{po}}}\,
\cos(rk\ell_{po}-rn_{po}\pi-\pi/4)
\end{equation}
and
\begin{equation}
\label{do}
\rho^{osc}_{do}(k)={\sum_{do}}^\prime\frac{\ell_{do}}{\pi}\frac{\mathcal{D}}{\sqrt{8\pi k \ell_{do}}}\,\cos(k\ell_{do}-n_{do}\pi-3\pi/4)
\end{equation}
where $\sum'$ denotes a sum over primitive periodic (diffractive) orbits of length $\ell_{po}$ ($\ell_{do}$) and number of bounces $n_{po}$ ($n_{do}$), $r$ is the number of repetitions. In formula (\ref{do}) only leading order one-scattering events are included, repetitions or concatenations of primitive orbits of order $\nu$ being of order $\mathcal{O}(k^{-\nu/2})$ \cite{Pavloff1995}.

Note that the contribution of DOs is subdominant with respect to POs due to different $k$-dependences. Nevertheless, the relevance of both POs and DOs is clearly illustrated through a weighted length spectrum \cite{Biswas93} which is obtained by calculating the so-called form factor $K(L)$, i.e. the Fourier transform of the spectral autocorrelation $C(\kappa)$ of $\rho^{osc}(k)$:

\begin{equation}
\label{correlation}
C(\kappa) = \langle  \rho^{osc}(k+\frac{\kappa}{2})  \rho^{osc}(k-\frac{\kappa}{2})
\rangle_{k}
\end{equation}

In practice, the local average over $k$ in (\ref{correlation}) can be written
 \begin{equation}
\label{moyenne}
\langle f(k)\rangle_k=\int\textrm{d}k'\,f(k')W_\sigma(k'-k)
\end{equation}
where the weighing function $W_\sigma$ is zero-centered and of typical width $\sigma$. In the following, we will use either Gaussian or Hanning weighing functions.

\begin{figure}[h]
\begin{center}
\includegraphics{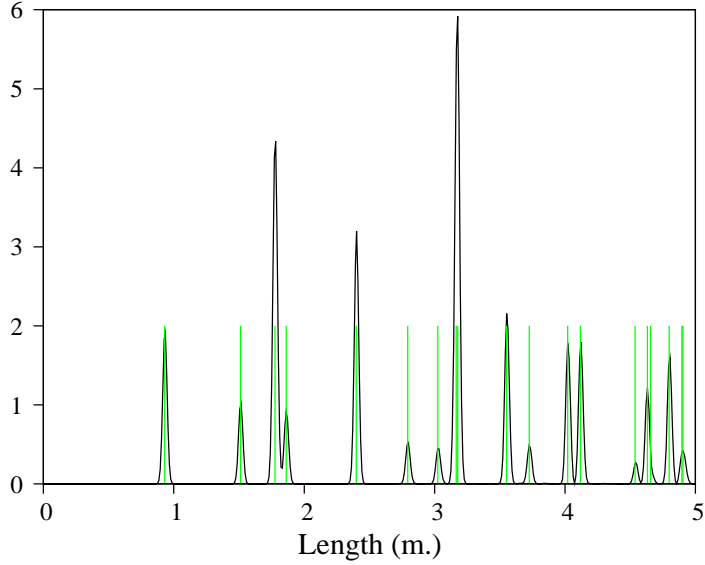}
\end{center}
\caption{Length spectrum computed in a rectangular cavity with a single point scatterer. Approximately 11000 resonances have been used. Green sticks indicate the lengths of the POs listed in Table \ref{tab1}.}
\label{LS}
\end{figure}

In the so-called \emph{diagonal approximation} \cite{Stoeckmann99}, the following expression of the form factor is obtained \cite{LegMorWea1997}:

\begin{equation}\label{K}
K(L) \propto \sum_{j} | A_{j} |^2 \delta(L-\ell_{j})
\end{equation}
Ideally, the length spectrum appears as a series of delta peaks located at particular orbits lengths $\ell_{i}$ with real positive amplitudes (see Figure \ref{LS}). 

For practical reasons, both in numerics and in analyzing our experiments, we will rather use the cumulated density $N(k)=\int^k_0\textrm{d}k'\,\rho(k')$, which is a staircase function increasing of one unit at each eigenwavenumber. This integrated quantity enables one to evaluate the form factor (\ref{K}) more readily.

\section{Numerical length spectra}

To validate the possiblity of identifying the shortest DOs in the short-term
non-universal part of the length spectrum, we first investigate a numerical
model of a rectangular cavity with a point scatterer (hereafter called the
\emph{dressed cavity}). To numerically obtain all the eigenfrequencies in a given frequency range for the dressed cavity, we will closely follow the method first introduced by R. L. Weaver and D. Sornette in reference \cite{WeaSor95}. This method provides a clear physical insight of the way the scatterer can be characterized by its transition matrix and scattering cross section. However, following a criticism made by Shigehara and Cheon in \cite{ShigeharaCheon96}, we will  correct a pathology of the aforementioned method by restoring the orthogonality of the perturbed eigenfunctions.

In 2D billiards, in the presence of a point scatterer at position $\vec{s}$, it is possible to write the total field $\Psi$ at the position $\vec{r}$, in the vicinity of $\vec{s}$, as the superposition of incoming and outgoing waves:

\begin{equation}
\label{Psi}
\Psi(\vec{r})\propto\left[\frac{1}{2}H_{0}^{(2)}(k|\vec{r}-\vec{s}\,|)+\left(\frac{1}{2}-\ir\frac{t}{4}\right)H_{0}^{(1)}(k|\vec{r}-\vec{s}\,|)\right]
\end{equation}
where $H_0^{(1,2)}$ are the Hankel functions of first and second kinds. The parameter $t(k)$ defines the ratio of the strength of the incident field at the scatterer to the
strength of  the outgoing field in the vicinity of the scatterer. In terms of this parameter the scattering cross section (here a length) is given by $|t|^{2}/4k$. From the expression (\ref{Psi}) for $\Psi$, energy conservation implies $\mid \frac{1}{2}\mid^{2}=\mid \frac{1}{2}-\ir\frac{t}{4}\mid^2$ yielding $t=(\alpha+\ir/4)^{-1}$, $\alpha$ being a real parameter. 

In the appendix, we show how a specific $k$-dependence of $\alpha$ can be established to ensure the orthogonality of the perturbed eigenfunctions. We thus reconcile the seemingly contradictory approaches of Refs \cite{WeaSor95} and \cite{ShigeharaCheon96}. The resulting expression for $t$ (equation (\ref{diffcoeff})) is the same as the diffraction constant in equation (\ref{DiffConst}).

Note that, apart from a logarithmic correction, the scattering cross section essentially scales as the wavelength, thus making the scatterer equally efficient at all frequencies, which is not the behavior of finite size scatterers as used in our experiments. However, in practice, we will always be in the limiting case $ka \ll 1$. The resonances of the dressed cavity correspond to the poles of the transition matrix (\ref{taumatrix}). Practically, the ensuing characteristic equation is solved through a convergence accelerating procedure described in \cite{WeaSor95}, by equating $\alpha (k) = - \left[ \, \ln (ka/2) + \gamma \, \right] / (2 \pi)$ to the function $g(k)$:

\begin{equation}
\label{gdek}
g(k)=\sum_{n} \Psi^{2}_{n}(\vec{s})\frac{4k^{6}}{k^{8}-(k_{n}^{_{(0)}})^8} - \frac{1}{4}
\end{equation}

Here, the $\Psi_{n}$'s  are the solution of the Helmholtz equation and the $k_{n}^{_{(0)}}$'s are the associated eigenwavenumbers in the empty cavity. The problem of calculating the eigenwavenumbers in the presence of the point scatterer is thus solved by finding the zeroes of $g(k)-\alpha(k)$. Note that $g(k)-\alpha(k)$ has singularities for each $k_{n}^{_{(0)}}$ and all its zeroes, hereafter denoted $k_i$, must lie between two consecutive eigenvalues of the empty cavity. 
This analytical approach allows us to deal with very large frequency ranges. 

In practice, we obtain the form factor and the length spectrum by calculating the Fourier transform of the difference $N(k)- \overline{N}(k)$, where the average behavior  $\overline{N}(k)$ is given by Weyl's formula:

\begin{equation}\label{weyl}
\overline{N}(k) = \frac{\mathcal{A}}{4\pi}k^2 - \frac{\mathcal{L}}{4\pi}k + \frac{1}{4}
\end{equation}
Here $\mathcal{A}$ is the total area of the cavity and $\mathcal{L}$ its perimeter.

\begin{figure}[h]
\begin{center}
\includegraphics{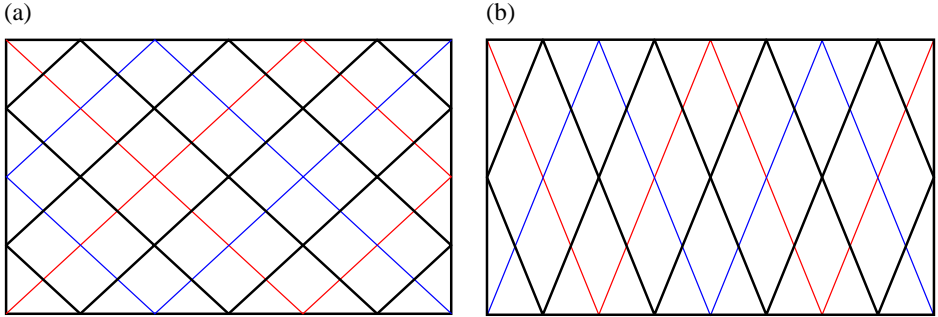}
\end{center}
\caption{Periodic orbits representation with indices (a) (2,3) (b) (1,4).  Continuous deformations of the orbit shown in thick black lines lead to the orbits shown in red or blue lines}
\label{fig2}
\end{figure}

An example of a length spectrum corresponding to a rectangular cavity with a single point scatterer is given in Figure \ref{LS}. Here, the dimensions of the cavity in which we have performed our calculations are those of the actual cavity we use in our experiments : $L_x=0.7562\,$m and $L_y=0.4656\,$m,  the scatterer is located at $x=46.5\,$cm and $y=20\,$cm (positions are measured from the upper left corner of the cavity sketched in Figures \ref{fig2} and \ref{fig4}), and the value of $a=3\,$mm corresponds to the radius of the actual cylindrical scatterer. A large number of resonances (approximately 11000) have been used so that the length resolution is excellent. To illustrate that such a length spectrum still is dominated by the POs of the empty cavity, it is shown in Figure \ref{LS} with an amplitude scale such that the contributions of the DOs are too small to be seen.

In a rectangular cavity of side lengths $L_x$ and $L_y$, the POs are identified by two indices $n$ and $m$ indicating the number of steps on a rectangular lattice of basis $(2L_x,2L_y)$ (see reference \cite{ Jain94}). They form families of continuously deformable orbits with the same length $\ell=2(n^2L_x^2+m^2L_y^2)^{1/2}$. For the families shown in Figure \ref{fig2}, the most symmetric representative is displayed, as well as the result of continuous deformations (red or blue) yielding self-retracing orbits hitting two corners. Note that, in polygonal billiards, diffraction occurs only at vertices where the angle is not a submultiple of $\pi$ \cite{Pavloff1995}. POs are easily identified  on the length spectrum shown in Figure \ref{LS} with the help of Table \ref{tab1} which gives the correspondence between indices and lengths for all lengths shorter than 5 meters. At first sight, it could even seem that no other contribution is to be seen as if the DOs were absent from it. Somehow, it could even be expected since no long range correlations are observed in the frequency spectrum thereby indicating that if the DOs should contribute, especially at short lengths, they should in a negligible way.

\begin{table}[h]
\caption{Indices and lengths (in meter) of the periodic orbits shown in Fig. 1}
\begin{center}
\begin{tabular}{cl@{ \quad }cl@{ \quad }cl@{ \quad }cl}
\hline
$(n,m)$ & length&$(n,m)$ & length & $(n,m)$ & length&$(n,m)$ & length\\
\hline
(0,1)Ê& 0.931 & (0,3) & 2.794 & (0,4) & 3.725 & (0,5) & 4.656\\
(1,0) & 1.512 & (2,0) & 3.025 & (1,4) & 4.020 & (2,4) & 4.798\\
(1,1) & 1.776 & (2,1) & 3.165  & (2,3) & 4.118 & (1,5) & 4.896\\
(0,2) & 1.862 &   (1,3) & 3.177 & (3,0) & 4.537  & (3,2) & 4.905\\ 
(1,2) & 2.399 & (2,2) & 3.552  & (3,1) & 4.632 & \multicolumn{2}{l}{} \\
\hline
\end{tabular}
\end{center}
\label{tab1}
\end{table}

\begin{figure}[h]
\begin{center}
\includegraphics{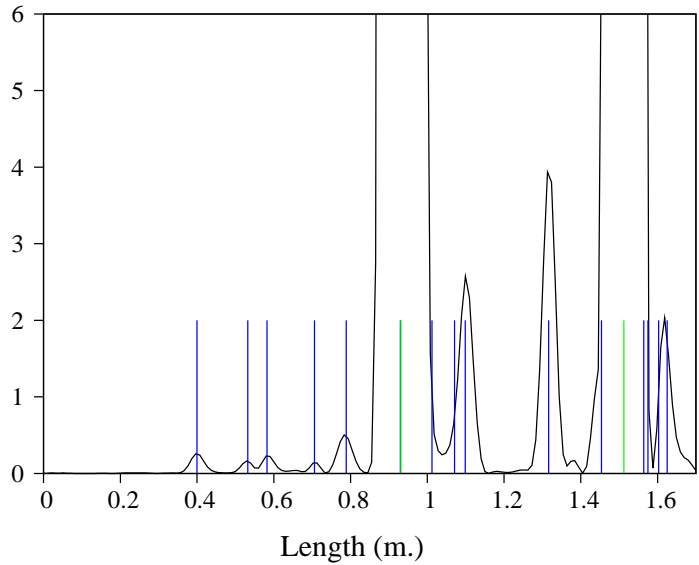}
\end{center}
\caption{Zoom ($\times 10^4$) of the length spectrum shown in Figure \ref{LS} on the length range from $0$ to $1.6$\,m using an enlarged scale for the amplitude of the peaks : POs (green sticks), DOs (blue sticks)}
\label{LSZ}
\end{figure}

\begin{figure}[h]
\begin{center}
\includegraphics{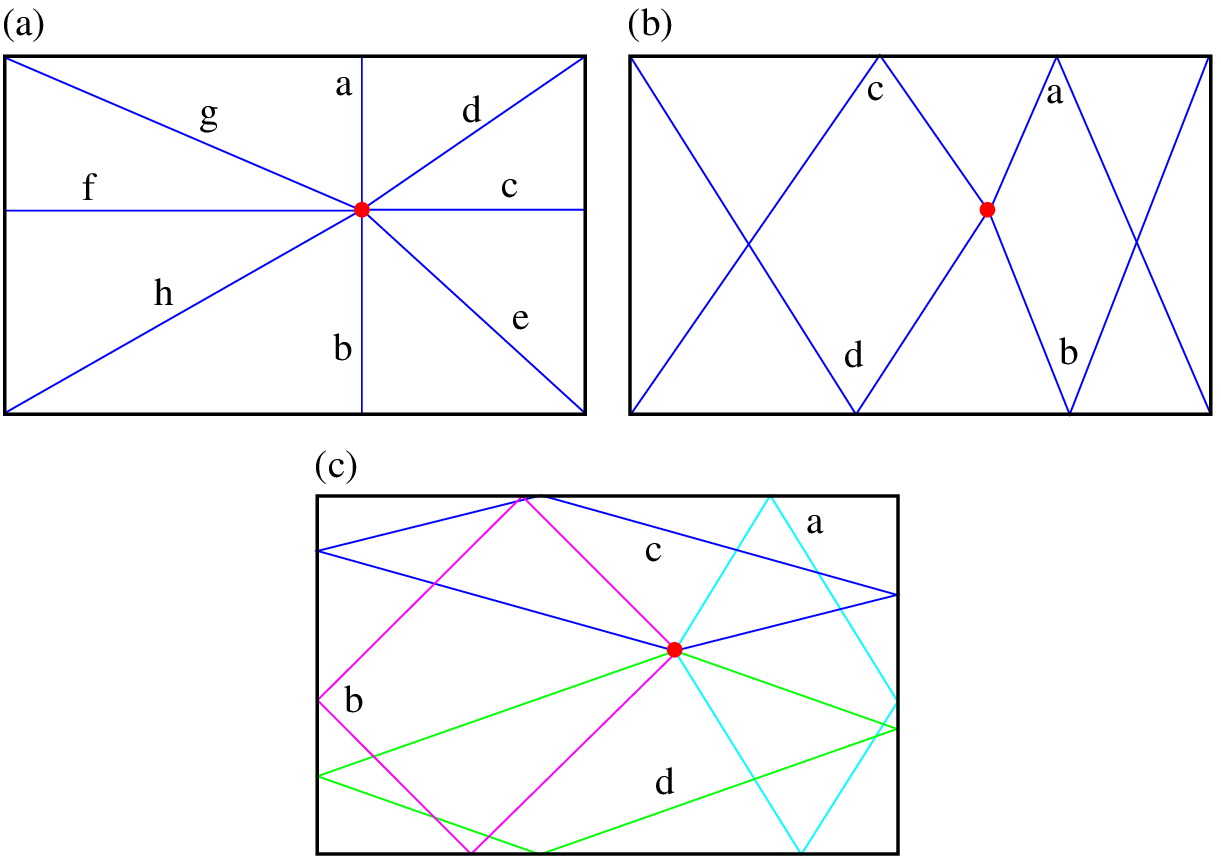}
\end{center}
\caption{Three different kinds of one-diffraction event  diffractive orbits associated to the position of the scatterer used numerically and experimentally:  (a) direct self-retracing DOs (1st kind), (b) self-retracing DOs (2nd kind), (c) 3-bounce DOs (3rd kind).}
\label{fig4}
\end{figure}

This is indeed what one can observe by closely inspecting a typical length spectrum for lengths smaller or of the order of the size of the cavity in the presence of a single point scatterer. In Figure \ref{LSZ}, the contributions of the DOs are displayed on the length range from $0$ to $1.6$\,m using an enlarged scale for the amplitude of the peaks. Sticks with different colors indicate the lengths of the DOs (blue) and POs (green) within this range. 

The shortest relevant lengths are those of the elementary self-retracing DOs starting from the scatterer and bouncing once on one of the sides of the rectangle as shown in Figure \ref{fig4}(a). In Figure \ref{fig4}(b), some of the shortest self-retracing DOs with one scattering event, two bounces on one side and one bounce in a corner are drawn. In Figure \ref{fig4}(c), DOs with one scattering event and three bounces are drawn. All these DOs are listed with their corresponding lengths in Table \ref{tab2}.

\begin{table}[h]
\caption{Lengths (in meter) of the diffractive orbits shown in Figure 4; 1st, 2nd and 3rd kinds refer to DOs depicted in (a), (b) and (c) respectively.}
\begin{center}
\begin{tabular}{l @{ : } l @{ \quad } l @{ : } l @{ \quad } l @{ : } l @{ \quad } l @{ : } l}
\hline
\multicolumn{4}{c@{ \quad }}{1st kind} & \multicolumn{2}{c@{ \quad }}{2nd kind} & 
\multicolumn{2}{c}{3rd kind}\\
\hline
a  &  0.400 & e  &  0.789 & a  &  1.454 & a  &  1.099\\
b  &  0.532 & f  &  0.930 & b  &  1.575 & b  &  1.316\\
c  &  0.582 & g  &  1.012 & c  &  1.625 & c  &  1.564\\
d  &  0.706 & h  &  1.071 & d  &  1.734 & d  &  1.603\\
\hline
\end{tabular}
\end{center}
\label{tab2}
\end{table}

\begin{figure}[h]
\begin{center}
\includegraphics[]{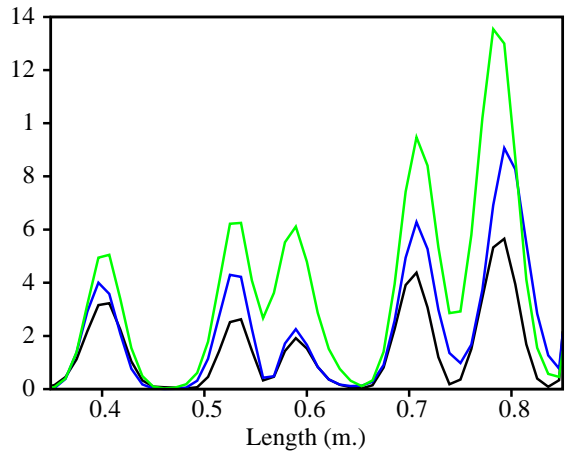}
\end{center}
\caption{Scaling law in $1/k$ of the amplitudes of the peaks associated to the DOs. Gaussian normalized frequency windows with equal variances (2\,GHz) centered on 10\,GHz (green), 15\,GHz (blue) and 20\,GHz (black) were used. An excellent agreement with the prediction of formula (\ref{do}) is observed.}
\label{scaling}
\end{figure}

In Figure \ref{scaling}, we have used different Gaussian normalized frequency windows with variance $\sigma=2\,$GHz centered on 10, 15 and 20\,GHz to illustrate the $1/k$-dependence of the squared amplitudes associated to the DOs (see formula (\ref{do})).  An excellent agreement with the prediction of formula (\ref{do}) is observed and validates the semiclassical approach.  We also have observed the $k$-dependence expected for POs (see formula (\ref{po})), but we will not present it here.

In this section we have used a model for a point-scatterer in a cavity to validate the possibility of revealing the presence of DOs in the short-term non-universal part of the length spectrum. 

\section{Experimental length spectra}

\begin{figure}[h]
\begin{center}
\includegraphics{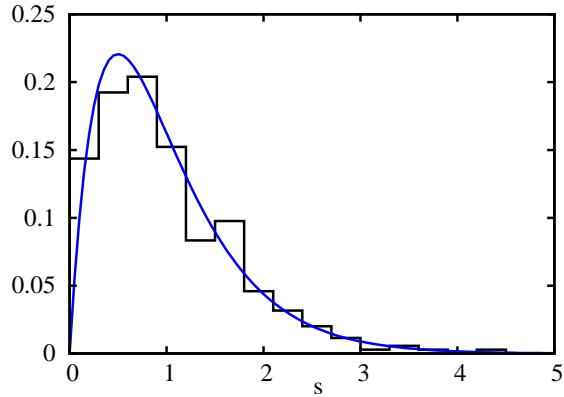}
\end{center}
\caption{Experimental histogram of P(s) in the dressed cavity; Semi-Poisson law: $P(s)=4s\textrm{e}^{-2s}$ (blue curve).}
\label{P(S)}
\end{figure}

In our experiments, the frequency spectra are determined from transmission signals measured in a 2D microwave cavity operated at frequencies ranging from 500\,MHz to 5\,GHz. The rectangular cavity is composed of two copper plates sandwiching a copper rectangular frame of thickness $5$\,mm and of width $2$\,cm. The cavity may thus be viewed as the slice of a rectangular waveguide closed at both ends, with perimeter $\mathcal{L}=2.446\,$m and area $\mathcal{A}=0.3528\,$m$^2$. The quality of copper is OFHC to reduce ohmic losses.  Due to its height of 5\,mm (smaller than half the smallest wavelength used), this cavity only admits transverse magnetic two-dimensional modes of order 0. Through one of the copper plates, ten antennas are introduced with positions determined at random. Optimal coupling was obtained by fixing their penetration length inside the cavity at 2\,mm. The details about transmission measurements are described in reference  \cite{Jerome_2004}. The scatterer is a small copper cylinder of radius 3\,mm much smaller than the smaller wavelength of the order of 5 \,cm. For the results presented here, we used three different couples of antennas to be sure not to miss any resonance frequency in the \emph{dressed} cavity in the range mentioned above (nearly 300 resonances were measured for each position of the scatterer). The cumulated density number $N(k)$ we could deduce from these measurements enabled us to verify that a small level repulsion can be observed as exemplified by the nearest-spacing distribution $P(s)$ whose histogram is shown in Figure \ref{P(S)}. The nearest-spacings are as usually obtained by building the sequence of normalized spacings $s_i = \overline{N}(k_{i+1}) - \overline{N}(k_{i})$, whose average is unity. A comparison is shown with the distribution associated to a semi-Poisson sequence as a guideline. In fact, Bogomolny \emph{et al.} have shown that the statistics of the singular billiard as the one discussed in the previous section, is intermediate between Poisson (uncorrelated spectrum) and semi-Poisson (short range repulsion) \cite{BogGerSch2001}.

\begin{figure}[h]
\begin{center}
\includegraphics[]{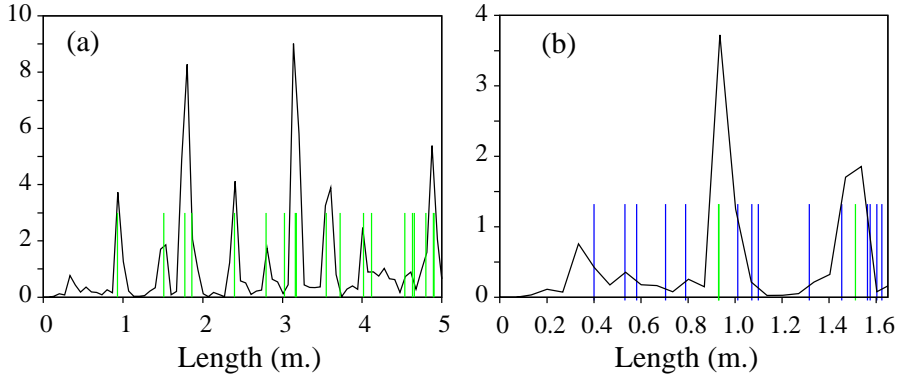}
\end{center}
\caption{Experimental length spectrum computed from 300 actual resonances measured from 500\,MHz to 5\,GHz. The position of the center of the scatterer is the same as in the numerical results of the previous section. (a) length range from $0$ to $5$\,m, POs indicated by green sticks; (b) length range from $0$ to $1.6$\,m, POs indicated by green sticks, DOs by blue sticks.}
\label{ExpLS}
\end{figure}

Once the frequencies of all the resonances in the range mentioned above have been determined with a sufficient precision (better than one tenth of the local spacing), we build an average cumulated density through a polynomial fitting of order two (consistent with Weyl's formula). Hence, an \emph{experimental} length spectrum is evaluated by using the total frequency range at our disposal (here, the weighing function of formula (\ref{moyenne}) is therefore simply a Hanning window) for the FFT.
Figure \ref{ExpLS} shows such a length spectrum on a scale where the shortest POs can clearly be identified. Note however the poor resolution compared to the \emph{analytical} length spectrum of Figures \ref{LSZ} or \ref{scaling}. This is entirely due to the reduced frequency range we were compelled to use. Indeed,  above 5\,GHz, the ohmic losses are so important that the overlap of neighboring resonances prevents from properly extracting all the resonance frequencies with the required precision. If one wishes to use frequencies above this limit, the resolution of the length spectrum is not significantly improved whereas the risk of missing many levels is rapidly increasing, thereby making the appearance of spurious peaks in the length spectrum more probable. These are current limitations when one requires high precision spectra in such microwave experiments at room temperature. Within this frequency range the resolution is sufficient to identify the shortest (and relevant) DOs, even when only one couple of antennas is used to extract the frequency spectrum.

\begin{figure}[h]
\begin{center}
\includegraphics[]{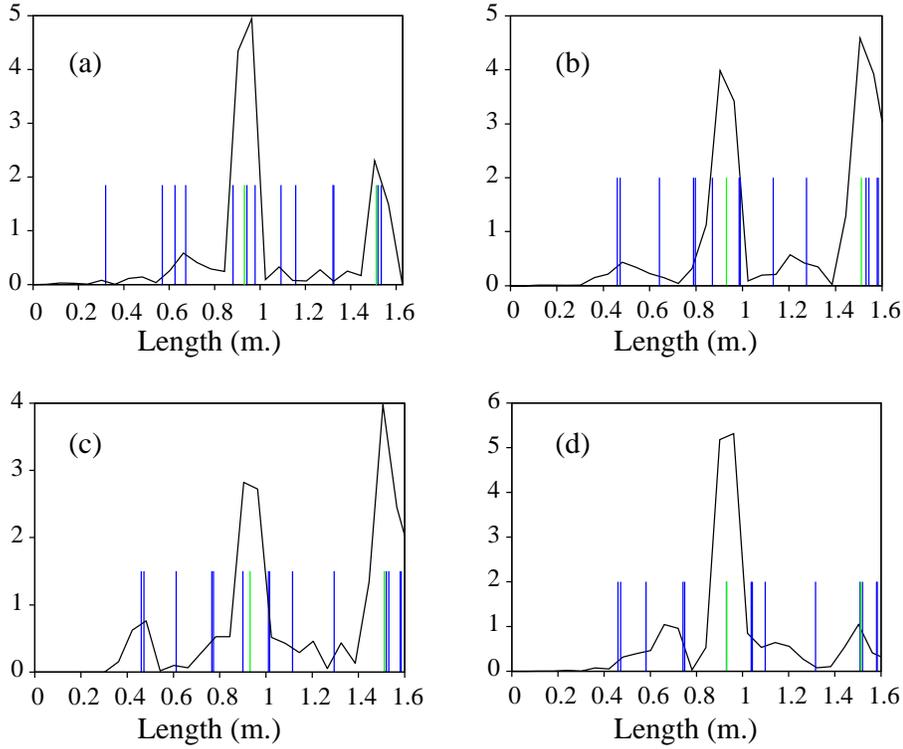}
\end{center}
\caption{Experimental length spectrum for 4 different positions of the scatterer measured from the upper left corner of the rectangular cavity: (a) $x=28.5\,$cm, $y=13\,$cm, (b) $x=43.5\,$cm, $y=23\,$ cm, (c) $x=45\,$cm, $y=23\,$cm, (d) $x=46,5\,$cm, $y=23\,$cm. The corresponding DOs of the three kinds are depicted by blue sticks. POs are indicated by green sticks.}
\label{ExpLSbis}
\end{figure}

As in the analytical part, one can note the presence of peaks not centered at the lengths of POs. The important difference with the analytical or numerical approach is that the peaks related to the DOs have amplitudes only one order smaller than those of the POs. This can be accounted by the finite size of the actual scatterer leading to a different $k$-dependence of the amplitudes associated to the DOs. 
In addition, we have also checked that the displacements of the scatterer within the cavity can be \emph{tracked} by properly monitoring the lengths of the peaks associated to the shortest DOs contributing to the \emph{experimental} length spectrum. This is illustrated in Figure \ref{ExpLSbis} where only one couple of antennas was used to extract the frequency spectrum for four different positions of the scatterer. This could constitute the basis of a novel non-destructive way of detecting the motion of a defect.

\section{Conclusion}

In the present paper, we have shown how a semiclassical approach of spectral statistics, usually based on periodic orbits in chaotic cavities, can be extended to analyze experiments in a 2D rectangular microwave cavity with a small scatterer by including diffractive orbits. More specifically, we have illustrated, through a model of cavity with a point-scatterer, how DOs can be clearly identified in the so-called \emph{length spectrum} of the cavity. This length spectrum could be retrieved through the evaluation of the two-point spectral correlations deduced from the cumulated spectral density of around 300 resonances measured in the frequency range [500\,MHz, 5\,GHz]. To our knowledge, this is the first experimental evidence of the unambiguous role played by diffractive orbits in the spectrum of an actual cavity. This is also a clear illustration of how a semi-classical analysis can be used to reveal non-universal spectral features of a complex cavity. Our findings show that the length spectrum can be used to locate the scatterer, or any small defect, and, eventually, to track a moving defect.

\section*{Acknowledgments}
We wish to thank Charles Poli for the numerical implementation of equation (\ref{gdek}). 
F. M. wishes to acknowledge enlightening discussions with Niels Sondergaard during the gathering \emph{Chaotic and Random Wave Scattering in Quantum Mechanics and Acoustics} held in Cuernavaca (Mexico, 2005). D. L. acknowledges financial support from DGA/CNRS Grant n. 2004487.
We especially wish to acknowledge the support of the Institute of Physics and of the Deutsche
Physikalische Gesellschaft in covering the publication costs.

\appendix
\section{Orthogonality of the perturbed eigenfunctions}

We will use the following notations: $G_{0}$ is the Green function of the bare cavity, $\Psi_{n}$ are the eigenfunctions and  $E_{n}$ the corresponding eigenenergies; $G$ is the perturbed Green function of the dressed cavity and $\Phi_{n}$ denotes the eigenfunctions with $z_{n}$ the corresponding eigenvalues. The two Green functions are related by:
\begin{equation}\label{Greenpert}
G \left( \vec{r},\vec{r}\,';z
\right)=G_{0}\left(\vec{r},\vec{r}\,';z\right)+G_{0}\left(\vec{r},\vec{s};z\right)\tau\left(\vec{s};z\right)G_{0}\left(\vec{s},\vec{r}\,';z\right)
\end{equation}
where the transition matrix $\tau$ is given by
\begin{equation}
\label{taumatrix}
\tau(\vec{s};z)=\left[t^{-1}-f\left(\vec{s};z\right)\right]^{-1}
\end{equation}
with $t^{-1}=\alpha+\ir/4$ (see formula (12) of reference \cite{WeaSor95}) for some real $\alpha$.

The resonances of the perturbed cavity correspond to the zeroes of $\left[t^{-1}-f\left(\vec{s};z\right)\right]$ where
\begin{eqnarray}
\label{f(b)}
f\left(\vec{s};z\right)&=&\lim_{\vec{r}\to
\vec{s}}\left[G_{0}\left(\vec{r},\vec{s};z\right)+\frac{\ir}{4}H_{0}^{\left(1\right)}\left(\sqrt{z}\|\vec{r}-\vec{s}\|\right)\right]\nonumber\\
&=&\frac{\ir}{4}+\lim_{\vec{r}\to
\vec{s}}\left[G_{0}\left(\vec{r},\vec{s};z\right)-\frac{1}{2\pi}\left(\gamma+\ln\frac{\sqrt{z}\|\vec{r}-\vec{s}\|}{2}\right)\right]
\end{eqnarray}
One deduces the $\Phi_{n}$ through the residues of $G$ at $z_{n}$
\begin{equation}
\lim_{z \to z_{n}}
\left[G\left(\vec{r},\vec{s};z\right)\left(z-z_{n}\right)\right]
=
\Phi_{n}\left(\vec{r}\right)\Phi_{n}\left(\vec{s}\right)
\end{equation}
Thus, according to (\ref{Greenpert}) (see \cite{ShigeharaCheon96}),  one has,
\begin{equation}
\Phi_{n}\left(\vec{r}\right) = N_{n}G_{0}\left(\vec{r},\vec{s};z_{n}\right)
\end{equation}
where
\begin{equation}
N_{n}^{2}=\left(\sum_{\nu}
\frac{\Psi_{\nu}^{2}\left(\vec{s}\right)}{\left(z_{n}-E_{\nu}\right)^2}\right)^{-1}
\end{equation}
One wishes to calculate
$\int\Psi_{m}\left(\vec{r}\right)\Psi_{n}\left(\vec{r}\right)\dr\vec{r}$. Using
\begin{equation}
G_{0}\left(\vec{r},\vec{s};z\right)=\sum_{\nu}\frac{\Psi_{\nu}\left(\vec{r}\right)\Psi_{\nu}\left(\vec{s}\right)}{z-E_{\nu}}
\end{equation}
one gets,
\begin{eqnarray}
&\int&\Phi_{m}\left(\vec{r}\right)\Phi_{n}\left(\vec{r}\right)\dr\vec{r}\nonumber\\
&=&N_{m}N_{n}\sum_{\nu,\mu}\frac{\Psi_{\nu}\left(\vec{s}\right)\Psi_{\mu}\left(\vec{s}\right)}{\left(z_{m}-E_{\nu}\right)
\left(z_{n}-E_{\mu}\right)}\int\Psi_{\nu}\left(\vec{r}\right)\Psi_{\mu}\left(\vec{r}\right)\dr\vec{r}\nonumber\\
&=&N_{m}N_{n}\sum_{\nu}\frac{\Psi_{\nu}^{2}\left(\vec{s}\right)}{\left(z_{m}-E_{\nu}\right)\left(z_{n}-E_{\nu}\right)}
\end{eqnarray}
where
$\int\Psi_{\nu}\left(\vec{r}\right)\Psi_{\mu}\left(\vec{r}\right)\dr\vec{r}=\delta_{\nu\mu}$ was used. If $m=n$, one finds:
\begin{equation}
\int\Phi_{n}^{2}\left(\vec{r}\right)\dr\vec{r}=N_{n}^{2}\sum_{\nu}\frac{\Psi_{\nu}^{2}\left(\vec{s}\right)}{\left(z_{n}-E_{\nu}\right)^{2}}
\end{equation}
\noindent
If $n\neq m$, one calculates
\begin{eqnarray}
f\left(\vec{s};z_{n}\right)&-&f\left(\vec{s};z_{m}\right)\nonumber\\
&=&\lim_{\vec{r}\to
\vec{s}}\left[G_{0}\left(\vec{r},\vec{s};z_{n}\right)-G_{0}\left(\vec{r},\vec{s};z_{m}\right)-\frac{1}{4\pi}\ln\left(\frac{z_{n}}{z_{m}}\right)\right]\nonumber\\
&=&\sum_{\nu}\frac{\Psi_{\nu}^{2}\left(\vec{s}\right)}{\left(z_{n}-E_{\nu}\right)}-\sum_{\nu}\frac{\Psi_{\nu}^{2}\left(\vec{s}\right)}{\left(z_{m}-E_{\nu}\right)}-\frac{1}{4\pi}\ln\left(\frac{z_{n}}{z_{m}}\right)\nonumber\\
&=&\left(z_{m}-z_{n}\right)\sum_{\nu}\frac{\Psi_{\nu}^{2}\left(\vec{s}\right)}{\left(z_{n}-E_{\nu}\right)\left(z_{m}-E_{\nu}\right)}-\frac{1}{4\pi}\ln\left(\frac{z_{n}}{z_{m}}\right)
\end{eqnarray}
At the resonances of $G$,
$f\left(\vec{s}\right)=t^{-1}$ assumed to be constant, whence
$f\left(\vec{s},z_{n}\right)=f\left(\vec{s},z_{m}\right)$, thus implying, if one writes $z_n=k_n^2$ :
\begin{equation}
\sum_{\nu}\frac{\Psi_{\nu}^{2}\left(\vec{s}\right)}{\left(z_{n}-E_{\nu}\right)\left(z_{m}-E_{\nu}\right)}=\frac{1}{2\pi\left(k_{m}^2-k_{n}^2\right)}\ln\left(\frac{k_{n}}{k_{m}}\right)\neq
0
\end{equation}
and finally yielding
\begin{equation}
\int\Phi_{m}\left(\vec{r}\right)\Phi_{n}\left(\vec{r}\right)\dr\vec{r}=\frac{1}{2\pi
N_{m}N_{n}}\frac{\ln\left(k_{n}/k_{m}\right)}{k_{m}^{2}-k_{n}^{2}} \, .
\end{equation}
If one allows $\alpha$ to depend on $k$, then one can restore the orthogonality between the eigenfunctions $\Phi_{n}$ and $\Phi_{m}$ with $m\neq n$. Indeed, it suffices to write:
\begin{equation}
\alpha\left(k_{n}\right)-\alpha\left(k_{m}\right)=-\frac{1}{2\pi}\ln\left(\frac{k_n}{k_m}\right)
\end{equation}
or, equivalently,
\begin{equation}
\alpha(k)=-\frac{1}{2\pi}\ln\left(\frac{k}{k_{0}}\right)+\textrm{const}
\end{equation}
If one puts $k_{0}=2/a$ and $\textrm{const}=-\gamma/(2\pi)$, $\gamma$ being the Euler constant, then (see \cite{ExnerSeba96})
\begin{equation}
\label{diffcoeff}
t=\frac{2\pi}{-\ln\left(\frac{ka}{2}\right)-\gamma+\ir\frac{\pi}{2}}
\end{equation}

This corresponds to the asymptotic behavior of the diffraction coefficient of a cylindrical s-wave scatterer of radius $a$. Therefore, in the approach due to  R. L. Weaver and D. Sornette \cite{WeaSor95}, the relation (23) which defines the resonances should read:
\begin{equation}
-\frac{1}{2\pi}\ln\left(\frac{ka}{2}\right)=
\lim_{\vec{r}\to \vec{s}}
\{G_{0}\left(\vec{r},\vec{s};z\right)-\frac{1}{2\pi}\ln\left(\frac{k\|\vec{r}-\vec{s}\|}{2}\right)
\}
\end{equation}
thus restoring \emph{the orthogonality of the perturbed eigenfunctions as well as the unitarity of the time-evolution operator} \cite{ShigeharaCheon96}.

\end{document}